\documentstyle[preprint,aps,prc]{revtex}

\begin{document}

\draft
\tighten

\vspace{1cm}

\title{Comment on the paper: ``Feynman-Schwinger representation 
approach to nonperturbative physics'' by \c C. \c Savkl\i \, et al. 
\\[0mm] 
[Phys. Rev. C 60, 055210 (1999), hep-ph/9906211]}

\author{R.~Rosenfelder $^1$ and A.~W. ~Schreiber $^2$}

\vspace{0.4cm}

\address{$^1$ Paul Scherrer Institute, CH-5232 Villigen PSI, Switzerland\\
$^2$ Department of Physics and Mathematical Physics and
           Research Centre for the Subatomic Structure of Matter,
           University of Adelaide, Adelaide, S. A. 5005, Australia}

\maketitle

\begin{abstract}
\noindent
We point out that the scalar Wick-Cutkosky model (with $\chi^2 \phi$ 
interaction) used  in this study has been known for a long time to be 
unstable. However, the numerical results presented in this paper do not 
show any sign of this instability which casts some doubt on their 
reliability. We compare with the worldline variational approach.
\end{abstract}


Worldline methods (sometimes called the ``particle represention of field 
theory'') have experienced a revival in the last few years (see e.g. 
Ref. \cite{worldline}), both in
perturbative and nonperturbative studies.
In a recent paper, \c Savkl\i \, {\it et al.} \cite {STG} have applied 
what they call the ``Feynman-Schwinger representation'' to 
various field theoretic  
models \cite{remark1} . Among these they also discuss the theory of 
charged scalar 
particles $\chi$ of mass $m$ interacting through the exchange of a neutral 
scalar particle $\phi$ of mass $\mu$ whose (Euclidean) Lagrangian is given
 by (Eq. (3.1))
\begin{equation}
{\cal L} \> = \> \chi^{\ast} \left [ m^2 - \partial^2 + g \phi \right ] \chi 
+ \frac{1}{2} \phi \left (\mu^2 - \partial^2 \right ) \phi \> .
\label{WC Lagrang}
\end{equation}
This is usually referred to as ``Wick-Cutkosky model'' \cite{WC} and has been 
studied extensively in the context of the bound-state problem in quantum field 
theory. In these studies self-energy corrections are generally omitted
and the exchange of neutral particles is restricted to be of the ladder type 
only.
However, it is well known that the full theory described by the 
Lagrangian (\ref{WC Lagrang}) is {\it unstable}. This is quite 
plausible already in a classical description by recognizing that 
the interaction term $g  \chi^{\ast} \chi \phi $ is equivalent to a 
$\Phi^3$-term whose potential is unbounded from below, but has been 
proven more rigorously by Baym \cite{Baym} nearly 40 years ago.
Note that this instability is also present in the {\it quenched} approximation
where closed particle loops are neglected. This can be easily seen, 
for example, from Eqs. (3.3) \cite{remark2} and (3.4) in Ref. \cite{STG}
which give the full two-particle propagator -- or any other $n$-point function --
in terms of
\begin{equation}
S(x,y) \> = \> \left < y \, \Bigg | \> \frac{1}{m^2 - \partial^2 + g \phi} 
\> \Bigg | \, x \right >
\end{equation} 
before the functional integration over the field $\phi$
has been performed. Although $m^2 -\partial^2$ is a positive definite operator 
it is obvious that for $ g \ne 0 $ there will be always (negative)
field configurations $\phi$ which lead to a vanishing of the denominator. 
If the singularity is properly treated, one therefore obtains an imaginary 
part of the euclidean $n$-point function \cite{McK} which for the propgator can 
be interpreted as width of the
metastable state \cite{RS2}. This happens irrespective whether the 
additional determinant in the functional integral over $\phi$ is set to a 
constant (quenched approximation) or taken fully into account \cite{remark3}.

\vspace{0.3cm}
The authors of Ref. \cite{STG} present Monte-Carlo results for the quenched
single-particle propagator (based on a discretized and Wick-rotated 
version of the Feynman-Schwinger representation) where the self-energy is 
the only mechanism to dress the bare propagator. Therefore their results 
should be sensitive to the above-mentioned deficiency of the 
Wick-Cutkosky model. However, Fig. 8 in their paper {\it shows no sign of the 
instability} over a wide range of coupling constants.
This casts serious doubts on the reliability of their numerical results and 
the claimed ``calculation of nonperturbative propagators''. We do not know
the reason for this failure: perhaps
it is due to the recipe used in Eq. (3.22) to suppress 
unwanted oscillations in the integral over the proper time $s$ or 
other numerical problems. Another possibility is that it results from
using (in the authors' words) ``a rather small
cutoff parameter'' $ \Lambda = 3 \mu$ in the Pauli-Villars 
{\it regularization}.
In any case, we believe that it is a serious failure and should be 
investigated in more detail.
In this context we also note that in Ref. \cite {STG} the necessary 
{\it renormalization} has not been performed, although, in principle,
it is straightforward in a super-renormalizable theory like the Wick-Cutkosky
model: the cutoff should be sent to infinity while keeping the physical mass
at its measured value. Without that a drastic cutoff-dependence of 
the results is inevitable.

\vspace{0.3cm}
The authors of Ref. \cite{STG} further present perturbative 
results for the self-energy of a single particle in the (quenched) 
Wick-Cutkosky model. 
Solving for the physical mass $M$ in the gap-like equation (3.40) they find 
that 
``the dressed mass $M$ decreases up to a critical value $g_{\rm crit}$ which 
occurs when the mass reaches ... $M_{\rm crit} = 0.094 $ GeV. For this simple 
case the critical coupling is given by
\begin{equation}
g_{\rm crit} \> = \> 22.2 \> \> \> {\rm GeV} \> .
\label{gcrit pert}
\end{equation}
For larger values of $g$ there are no real solutions, showing that the 
dressed particle is 
unstable''. As we have pointed out to the authors \cite{Ro} this critical 
coupling constant is a far cry from the one obtained in a worldline 
variational approach \cite{RS2} which gave  $ \> 
\alpha_{\rm crit} \> \equiv \> \bar g_{\rm crit}^2/(4 \pi M^2) \> = \>  
0.81 $ where $ 2 \bar g \equiv g$ (cf. Eq. (15) in Ref. \cite{RS1} with Eq.
(3.4) in the paper under discussion).
In this calculation, the physical mass $M$ was always kept 
at $0.939$ GeV and $\mu = 0.14$ GeV.
In contrast, Eq. (\ref{gcrit pert}) for the slightly different value 
$\mu = 0.15$  GeV leads to the totally unrealistic value 
$\alpha_{\rm crit} = (11.1/0.094)^2/(4 \pi) \simeq 1100 $
which only shows that perturbation theory cannot be applied in the 
nonperturbative regime.

\vspace{0.3cm}
In summary, we think that one of the few things one can learn from applying
nonperturbative methods to an unrealistic model like the Wick-Cutkosky model 
is whether the specific method is capable to
catch some genuine nonperturbative aspects. In the one-(heavy)-particle sector 
this is mainly the instability of that model \cite{remark4}.
Neither the perturbative calculation  nor the supposedly ``exact'' 
numerical Monte-Carlo calculation in Ref. \cite{STG} fare well in this respect.


\begin{thebibliography}{99}
\bibitem{remark2} Note that the power of the determinant is incorrect as
$S$ is already the inverse of the operator $m^2 - \partial^2 + g \phi$.
The same remark applies to Eq. (2.7).
\end{thebibliography}
\end{document}